\documentclass[12pt,preprint]{emulateapj}
\usepackage{graphicx}
\usepackage{amssymb,amsfonts,amsmath,amstext,amsgen,amsopn,amsxtra,indentfirst,times}

\begin{document}

\title{On the Existence of Shocks in Irradiated Exoplanetary Atmospheres}
\author{Kevin Heng\altaffilmark{1,2}}
\altaffiltext{1}{ETH Z\"{u}rich, Institute for Astronomy, Wolfgang-Pauli-Strasse 27, CH-8093, Z\"{u}rich, 
Switzerland}
\altaffiltext{2}{Zwicky Prize Fellow}

\begin{abstract}
Supersonic flows are expected to exist in the atmospheres of irradiated exoplanets, but the question of whether shocks develop lingers.  Specifically, it reduces to whether continuous flow in a closed loop may become supersonic and if some portions of the supersonic flow steepen into shocks.  We first demonstrate that continuous, supersonic flow may exist in two flavors: isentropic and non-isentropic, with shocks being included in the latter class of solutions.  Supersonic flow is a necessary but insufficient condition for shocks to develop.  The development of a shock requires the characteristics of neighboring points in a flow to intersect.  We demonstrate that the intersection of characteristics may be quantified via knowledge of the Mach number.  Finally, we examine 3D simulations of hot Jovian atmospheres and demonstrate that shock formation is expected to occur mostly on the dayside hemisphere, upstream of the substellar point, because the enhanced temperatures near the substellar point provide a natural pressure barrier for the returning flow.  Understanding the role of shocks in irradiated exoplanetary atmospheres is relevant to correctly modeling observables such as the peak offsets of infrared phase curves.
\end{abstract}


\section{Introduction}

Current astronomical techniques favor the detection of close-in, irradiated exoplanets that are believed to be tidally locked.  With temperatures reaching $\sim 1000$--3000 K, the flows in these atmospheres are expected to approach or exceed the local sound speed ($\sim 1$ km s$^{-1}$).  A lingering question in the astrophysical literature is whether these supersonic flows can develop into shocks and subsequently dissipate some of the kinetic energy of the flow as heat.  Specifically, the question reduces to whether continuous flow in a closed loop may become supersonic and if some portions of the supersonic flow steepen into shocks.

In the present Letter, we provide the answers to these questions by re-visiting a problem that is mathematically and physically similar: the design of wind tunnels and the de Laval nozzle.  In \S\ref{sect:delaval}, we demonstrate the possibility of supersonic flow in a closed-loop flow and state a simple criterion for determining if these flows develop shocks.  In \S\ref{sect:hj}, we apply our method to simulations of hot Jovian atmospheres.  The implications of our study are discussed in \S\ref{sect:discussion}.

\section{The de Laval Nozzle and Wind Tunnel Design}
\label{sect:delaval}

While much of this section is based on the monograph of \cite{lr57}, the re-derivation of these results elucidates the key concepts and assumptions made in stating our method, which we will later apply to hot Jovian atmospheres.

\subsection{Demonstrating the Possibility of Supersonic Flow}

In its most basic form, the de Laval nozzle consists essentially of a tube that continuously narrows to a minimum in its cross section, known as the ``throat", before widening again in a symmetric fashion.  The conservation of mass dictates that the mass flux $\dot{M} = \rho v A$ is constant, such that
\begin{equation}
\frac{dA}{A} + \frac{dv}{v} + \frac{d\rho}{\rho} = 0,
\end{equation}
where $A$ is the cross-sectional area of the nozzle, $v$ is the flow speed normal to the cross section and $\rho$ is the mass density of the fluid.  Invoking the conservation of momentum ($dP = \rho v dv$ with $P$ being the pressure) and the definition for the sound speed ($c_s^2 = \partial P/\partial\rho$), we obtain
\begin{equation}
\frac{d\rho}{\rho} = -{\cal M}^2 \frac{dv}{v},
\end{equation}
with ${\cal M} \equiv v/c_s$ being the Mach number of the flow.  When the flow is strictly subsonic (${\cal M} \ll 1$), it becomes virtually incompressible ($d\rho \approx 0$).  Substituting this expression into the equation for the conservation of mass yields
\begin{equation}
-\frac{dA}{A} = \frac{dv}{v} \left( 1 - {\cal M}^2 \right).
\label{eq:delaval_basic}
\end{equation}

When the flow is subsonic (${\cal M} < 1$), narrowing the nozzle ($dA < 0$) leads to an increase in the flow speed ($dv > 0$).  The flow is symmetric with respect to the throat.  However, if the flow is supersonic (${\cal M} > 1$), narrowing the nozzle leads to the somewhat counter-intuitive result that a \emph{decrease} in the flow speed ($dv < 0$) occurs.  The flow is now \emph{asymmetric} across the throat because of compressibility effects.  In the design of continuous wind tunnels there is often a region of narrowed cross section, downstream of the de Laval nozzle, known as a ``diffuser", which decelerates the flow and discharges it back into the return circuit.  When $dA=0$, we get $dv \ne 0$ only when ${\cal M}=1$.

\begin{figure}
\centering
\includegraphics[width=\columnwidth]{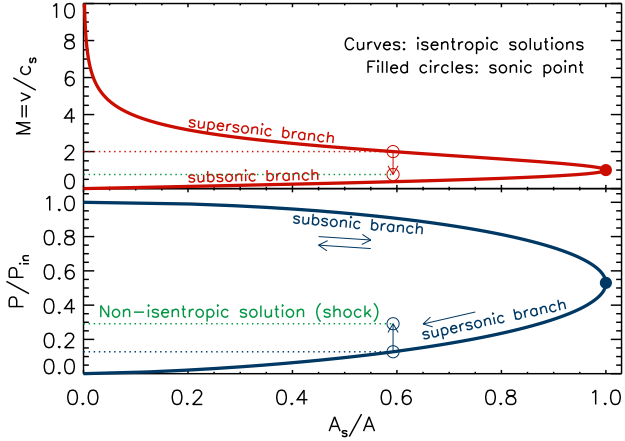}
\caption{Subsonic and supersonic solutions for the de Laval nozzle as functions of $A_s/A$, where $A$ is the cross-sectional area and $A_s$ is the cross-sectional area corresponding to the sound speed.  Top panel: Mach number.  Bottom panel: ratio of pressure to entry pressure.  The sonic point is marked by a filled circle.  The empty/unfilled circles indicate a possible, non-isentropic solution for a ${\cal M}=2$ shock.}
\vspace{0.2in}
\label{fig:delaval_solutions}
\end{figure}

The entry into the de Laval nozzle is described by a (small) velocity $v_{\rm in}$, a temperature $T_{\rm in}$, a pressure $P_{\rm in}$ and a mass density $\rho_{\rm in}$.  By enforcing the conservation of energy and assuming an ideal gas, we may transform the isentropic relations into ones for the pressure and density ratios,
\begin{equation}
\begin{split}
&\frac{P_{\rm in}}{P} = \left[ 1 + \frac{1}{2} \left( \gamma-1 \right) {\cal M}^2 \right]^{\gamma/\left(\gamma-1\right)},\\
&\frac{\rho_{\rm in}}{\rho} = \left[ 1 + \frac{1}{2} \left( \gamma-1 \right) {\cal M}^2 \right]^{1/\left(\gamma-1\right)},
\end{split}
\label{eq:prho_ratios}
\end{equation}
where $\gamma = 1 + 2/n_{\rm dof} = 7/5$ is the adiabatic gas index and $n_{\rm dof}=5$ is the number of degrees of freedom in the gas.  Note that $P_{\rm in} > P$ and $\rho_{\rm in} > \rho$.  By enforcing the conservation of mass, one obtains the cross-sectional area normalized by its sonic value,
\begin{equation}
\frac{A}{A_s} = \frac{1}{{\cal M}}\left\{ \frac{2}{\gamma+1} \left[ 1 + \frac{1}{2}\left( \gamma-1 \right){\cal M}^2 \right] \right\}^{\left(\gamma+1\right)/2\left(\gamma-1\right)}.
\label{eq:area_mach}
\end{equation}
By utilizing equation (\ref{eq:prho_ratios}), one can eliminate ${\cal M}$ in favor of $P/P_{\rm in}$,
\begin{equation}
\begin{split}
\frac{A}{A_s} =& \left( \frac{\gamma-1}{2} \right)^{1/2} \left( \frac{2}{\gamma+1} \right)^{\left(\gamma+1\right)/2\left(\gamma-1\right)} \\
&\times \left( \frac{P}{P_{\rm in}} \right)^{-1/\gamma} \left[ 1 - \left(\frac{P}{P_{\rm in}}\right)^{1-1/\gamma} \right]^{-1/2}.
\end{split}
\label{eq:area_pressure}
\end{equation}

Figure \ref{fig:delaval_solutions} shows ${\cal M}$ and $P/P_{\rm in}$ as functions of $A_s/A$ for $\gamma=7/5$.  The cross-sectional area $A=A(x)$ serves as a proxy for the actual distance $x$ along the nozzle.  There are generally two solution branches: subsonic and supersonic.  If a given nozzle has a throat with a cross-sectional area larger than $A_s$, then the solution remains in the subsonic branch and is symmetric about the throat.  If the throat has a cross-sectional area $A_s$, then the system develops an asymmetry, where the flow from the entry point to the throat is subsonic, while the flow from the throat to the exit point is supersonic.  The pressure drops monotonically from a value of $P_{\rm in}$ to $P_{\rm out}$.  If the flow remains isentropic, then the exit pressure $P_{\rm out}$ is determined by the supersonic solution to the $P/P_{\rm in}$ curve in Figure \ref{fig:delaval_solutions}.  Generally, isentropic solutions cannot exist between the two bracketing values of $P/P_{\rm in}$ for a given $A_s/A$.  However, \emph{non-isentropic} solutions may exist.  In the example shown, a Mach-2 shock occurs in the flow.  The existence of the shock creates entropy and re-compresses the flow,
\begin{equation}
\frac{P^\prime}{P_{\rm in}} = \frac{P}{P_{\rm in}} \left( \frac{2\gamma {\cal M}^2 -\gamma + 1}{\gamma+1} \right).
\end{equation}
Furthermore, the Mach number is reduced,
\begin{equation}
{\cal M}^\prime = \left[ \frac{\left(\gamma-1\right){\cal M}^2 + 2}{2\gamma {\cal M}^2 + 1 - \gamma} \right]^{1/2}.
\end{equation}
Post-shock quantities are indicated with a prime.

While our present study of the de Laval nozzle cannot be directly applied to exoplanetary atmospheres and thus does not yield any quantitative predictions, we do gain some heuristic insight:
\begin{itemize}

\item We expect supersonic flow to be able to develop in continuous flows inherent in exoplanetary atmospheres (without orography), consistent with the results of published 3D simulations (see \S\ref{subject:previous} for references);

\item In the de Laval nozzle, variations in the pressure, density and Mach number are induced by changes in the cross-sectional area.  In irradiated atmospheres, changes in the Mach number are induced by changes in temperature;

\item For close-in exoplanets that are expected to be tidally locked, the hemispheric (dayside-only) nature of the stellar irradiation \emph{naturally} introduces re-compression into the continuous flow from dayside to nightside.  Irradiated, tidally-locked exoplanets create their own pressure ``barrier" for the atmospheric flow to crash into, thus creating shocks.  We expect this pressure barrier to exist near the substellar point.

\end{itemize}

\subsection{Demonstrating the Existence of Shocks}

The question of whether a shock develops in a continuous flow can be reduced to analyzing how the Mach number of the flow changes along a given streamline.  Since mass, momentum and energy are conserved, the problem is mathematically identical to employing the established machinery of the Rankine-Hugoniot jump conditions.

Consider a flow in 2D, Cartesian coordinates.  Let the flow be described by the velocity components $v_x$ and $v_y$, such that $v^2 = v^2_x + v^2_y$.  Across an interface, the Mach number changes.  Let the angle between the flow and the interface be $\beta$ such that $\tan \beta = v_x/v_y$.  Downstream of the interface, the flow velocity is $v^\prime$ and the flow direction is deflected by an angle $\theta$ relative to $v$, yielding $\tan (\beta - \theta) = v^\prime_x/v^\prime_y$.  Denoting the upstream Mach number by ${\cal M} \equiv v/c_s$, the ratio of the downstream to the upstream mass density is
\begin{equation}
\frac{\rho^\prime}{\rho} = \frac{\left( \gamma+1 \right) {\cal M}^2 \sin^2\beta}{\left(\gamma-1\right) {\cal M}^2 \sin^2\beta + 2}.
\end{equation}
A relationship between $\beta$ and $\theta$ may be obtained by taking the ratio of $\tan\beta$ to $\tan(\beta-\theta)$ and invoking the jump condition for $\rho^\prime/\rho$,
\begin{equation}
{\cal M}^2 \sin^2\beta - 1 = \frac{\left( \gamma + 1 \right) {\cal M}^2 \sin\beta \sin\theta}{2 \cos\left(\beta - \theta\right)}.
\label{eq:angle_relation}
\end{equation}
For $\theta=0^\circ$, two corresponding values of $\beta$ exist.  The first is $\beta=90^\circ$, which describes a situation where the downstream flow is subsonic (e.g., for a strong shock).  The second occurs when
\begin{equation}
\beta_0 \equiv \sin^{-1}\left(1/{\cal M}\right),
\end{equation}
which describes a situation where the downstream flow is supersonic.

Supersonic flow is a necessary but insufficient condition for a shock to develop.  A wave with a constant propagation speed possesses the property that every point on the wave is translated by the same distance after a given time interval.  In the parameter space of distance versus time, a given point on the wave at different times is connected by a straight line known as a characteristic.  It is apparent that these characteristics never intersect.  In reality, the propagation speed also depends on the density of the fluid, implying that different points on the wave are translated by different distances over a given time interval.  In this case, the characteristics may intersect---if they do, a shock forms and converts part of the kinetic energy into heat.

At the equator of an irradiated, tidally-locked exoplanet, the flow is predominantly zonal (east-west).  For purely zonal flow, each characteristic is associated with its own value of the angle $\beta_0$ (sometimes known as the ``Mach angle").  When $\beta_0$ decreases, the characteristics diverge.  When $\beta_0$ increases, they converge.  Convergence is not necessary an indication of intersection---there needs to be enough \emph{room} for the characteristics to intersect.  On a sphere (with a radius $R$) where two points are separated by a small angular distance $\Delta \phi$, the horizontal distance $h$ over which intersection occurs is given by
\begin{equation}
\frac{h}{R} = \Delta \phi \left( \frac{1}{\tan\beta_{0,1}} - \frac{1}{\tan\beta_{0,2}} \right)^{-1},
\end{equation}
where the subscript ``2" refers to a point downstream of another point subscripted by ``1".

The method stated in this sub-section may be applied to local pairs of points in a global flow. 

\section{Application to Hot Jupiters}
\label{sect:hj}

\subsection{Analytical Models}

\begin{figure}
\centering
\includegraphics[width=\columnwidth]{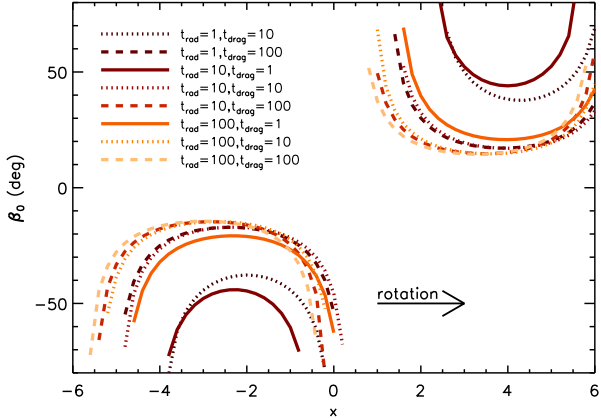}
\caption{Characteristic angles $\beta_0$, at the equator, computed using the Showman \& Polvani (2011) analytical model.  Shown are models with different values of the dimensionless radiative ($t_{\rm rad}$) and hydrodynamic ($t_{\rm drag}$) drag time scales.  For the model with $t_{\rm rad} = t_{\rm drag} = 1$, the dimensionless velocity (see text for definition) is always less than unity and thus $\beta_0$ is undefined.  The dimensionless distance $x$ is given in units of the Rossby length scale.}
\label{fig:sp11}
\end{figure}

We first apply the method described in the previous section to the 2D, linear, analytical, 1.5-layer, shallow water model of \cite{sp11}, which reproduces some of the flow features of hot Jovian photospheres.  There are several important limitations associated with the shallow water model.  The governing equation for the shallow water height derives from the condition of incompressibility.  There is no governing equation for temperature.  Both features imply that a sound speed cannot formally be defined.  However, a proxy for the sound speed is the gravity wave speed, which is defined in shallow water systems as $c_g = \sqrt{gH}$ where $g$ is the surface gravity and $H$ is the equilibrium water height.  When the flow velocity exceeds $c_g$, the analogues of shocks known as hydraulic jumps develop.  The flow velocity normalized by $c_g$ is then a proxy for the Mach number.

Stellar irradiation and radiative transfer are not explicitly considered in the shallow water model.  Rather, the effects of radiation are mimicked using a dimensionless radiative drag time scale $t_{\rm rad}$.  A dimensionless time scale $t_{\rm drag}$ mimics the effect of hydrodynamic drag.  Figure \ref{fig:sp11} shows calculations corresponding to the 9 models presented in Figure 3 of \cite{sp11}.  We have shown the characteristic angle $\beta=\beta_0$ only at the equator, since this simplifies our analysis.  Negative values of $\beta_0$ correspond to counter-rotating (rather than super-rotating) flow.  

First, we see that there are locations where the magnitude of $\beta_0$ increases and thus hydraulic jump formation may occur.  Second, the existence of these locations is robust to the strength of irradiation (through $t_{\rm rad}$) and the presence of drag (through $t_{\rm drag}$).  However, since we do not expect the shallow water model to completely reproduce the flow field computed from 3D simulations, we do not expect it to correctly mimic the locations of shock formation.

\subsection{3D Simulations}

\begin{figure}
\centering
\includegraphics[width=\columnwidth]{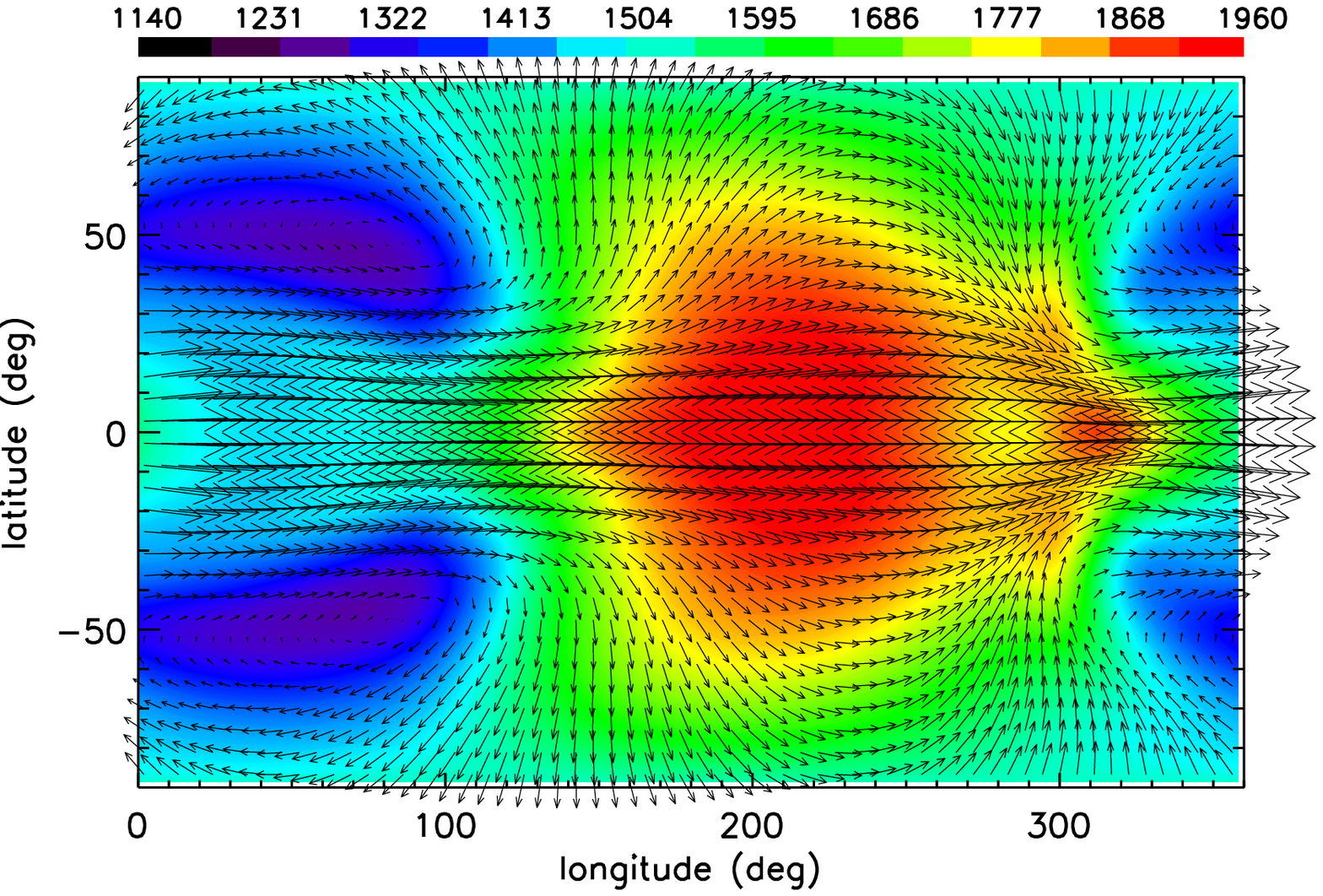}
\includegraphics[width=\columnwidth]{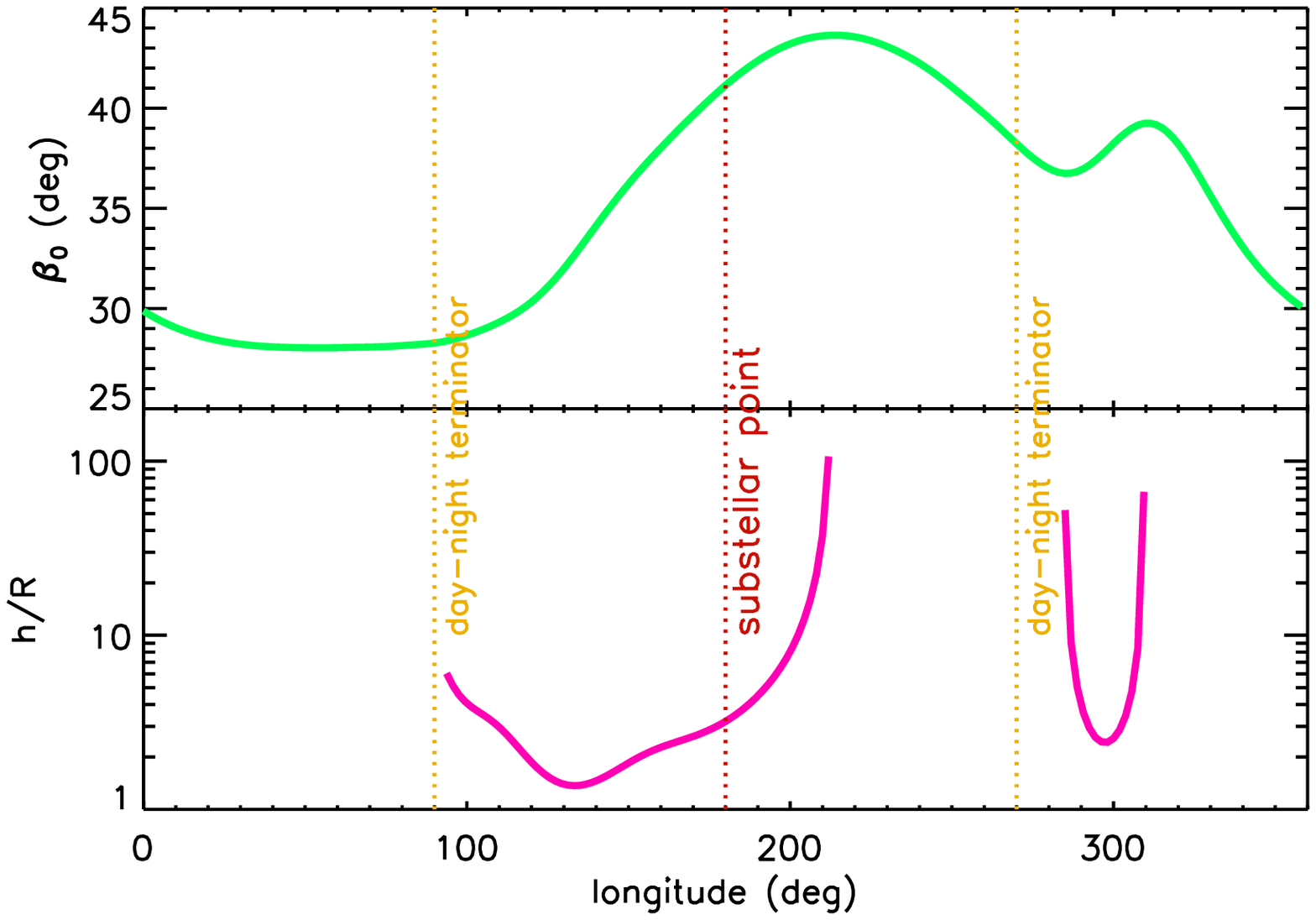}
\caption{Determining the locations where shocks form near the infrared photospheres of hot Jupiters.  The equilibrium temperature adopted in this model is about 1720 K (Model H).  Top panel: map of temperature (colors; in units of K) and velocity (arrows) at $P \approx 10$ mbar.  Middle and bottom panels: $\beta_0$ and $h/R$, at the equator, as functions of longitude.}
\label{fig:model_h}
\end{figure}

\begin{figure}
\centering
\includegraphics[width=\columnwidth]{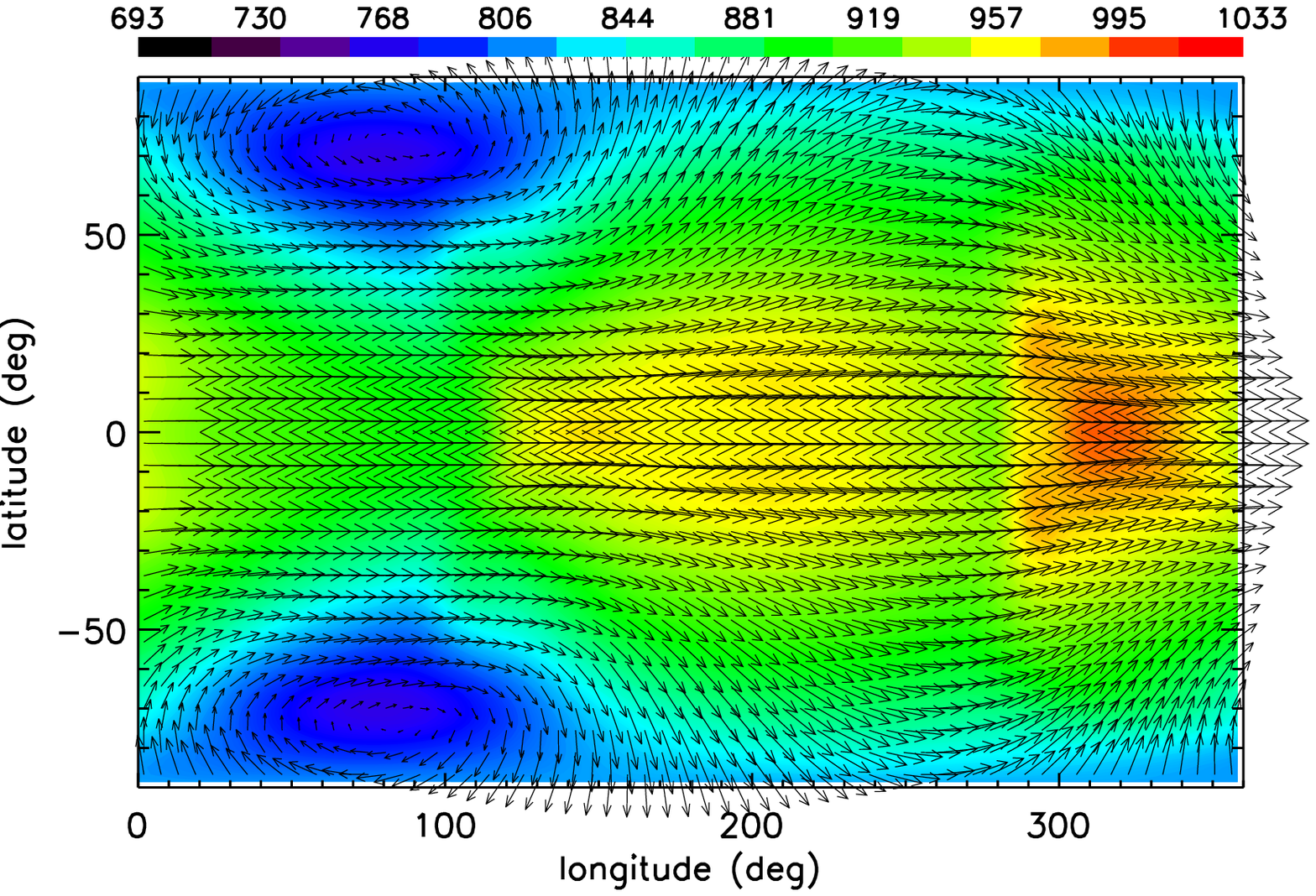}
\includegraphics[width=\columnwidth]{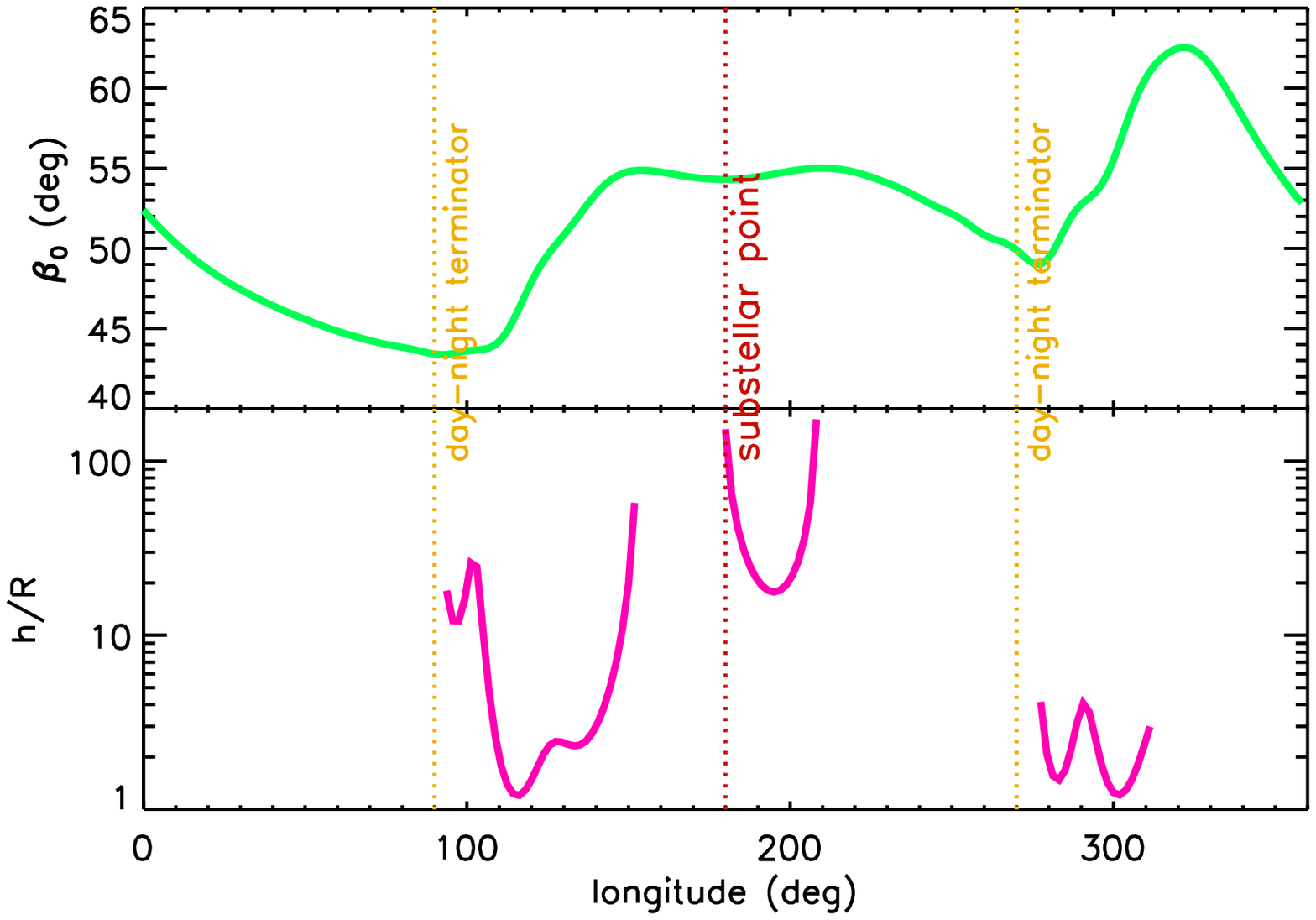}
\caption{Same as Figure \ref{fig:model_h}, but for an equilibrium temperature of about 970 K (Model W).}
\label{fig:model_w}
\end{figure}

We next apply our method to 3D simulations of atmospheric circulation of hot Jupiters, which solve for atmospheric dynamics and dual-band radiative transfer self-consistently \citep{hmp11,hfp11}.  Specifically, we first adopt Model H of \cite{php12} as a case study, which has an equilibrium temperature of $T_{\rm eq} \approx 1720$ K and has no temperature inversion present in its atmosphere ($\gamma_0=0.5$).  (See Tables 1 and 2 of \citealt{php12} for more details.)\footnote{We note a typographical error in Table 1 of \cite{php12}: $\tau_{\rm S_0}$ should have values of $5 \times 10^3$ and $2 \times 10^4$.}

The top panel of Figure \ref{fig:model_h} shows the temperature-velocity map near the infrared photosphere of our model hot Jupiter.  As witnessed in published 3D simulations of hot Jovian atmospheres, a global, chevron-shaped feature is centered about the equator, which is a mostly linear response arising from the interaction of the mean flow with standing Rossby and Kelvin waves \citep{sp11}.  

The middle panel of Figure \ref{fig:model_h} shows the characteristic angle $\beta_0$ versus longitude at the equator.  The corresponding Mach numbers are supersonic throughout.  In the direction of the zonal flow, $\beta_0$ increases with longitude largely on the dayside hemisphere.  At locations where shock formation has a chance of occurring, we estimate the values of $h/R$ (bottom panel of Figure \ref{fig:model_h}).  In the case of Model H, locations exist where $h/R \lesssim \pi/2$ implying that the intersection of characteristics, associated with points in the flow where $\beta_0$ is increasing, will occur.  We reach the same conclusion when we examine atmospheric layers at higher altitudes (lower pressures).  At pressures of $P \gtrsim 1$ bar, the equatorial Mach numbers become less than unity and the necessary condition for shock formation is unfulfilled.

In Figure \ref{fig:model_w}, we examine Model W of \cite{php12}, which has $T_{\rm eq} \approx 970$ K.  Our conclusions drawn from examining Model H carry over: shocks form mostly on the dayside hemisphere, where $\beta_0$ is an increasing function of longitude.  We also examine Models W and H with temperature inversions present ($\gamma_0=2$).  Besides minor quantitative differences between the results from these models and those presented in Figures \ref{fig:model_h} and \ref{fig:model_w}, we find no qualitative differences.  Models with lower equilibrium temperatures ($T_{\rm eq} \approx 540$--750 K; Models C, C1 and C2) do not develop supersonic flows.

In all of the cases examined, shocks---when they do form---are expected to form mainly \emph{upstream} of the substellar point.  \emph{The enhanced temperatures near the substellar point are creating a natural pressure barrier for the returning flow, from the nightside, to ``crash" into, thus creating conditions conducive to shock formation.}\footnote{We may gain further insight by recalling the classical situation of a ${\cal M} \gg 1$ flow impinging upon a physical barrier, which forces the downstream velocity to ultimately become zero.  In this case, an increasing $\beta_0$ is enforced by decreasing $v$.  In irradiated exoplanetary atmospheres, an increasing $\beta_0$ is enforced by increasing $T$.}  While we have focused on hot Jupiters, we expect the insight gained from the present study to apply to all types of hot exoplanets---Earth-like, Neptune-like or Jupiter-like.  The atmospheres of tidally-locked, irradiated exoplanets are akin to \emph{forced} wind tunnels.

\section{Discussion}
\label{sect:discussion}

\subsection{Relevance to Previous Work}
\label{subject:previous}

Previously published work on simulations of irradiated exoplanetary atmospheres have recorded supersonic flows when the temperature reaches $\sim 1000$ K.  Most of these studies utilize ``general circulation models" (GCMs), which are traditionally designed to model the terrestrial climate system and do not include an explicit treatment of shocks \citep{showman09,tc11,hmp11,hfp11,rm12}.  These simulations may describe supersonic flow, but do not correctly convert a fraction of the kinetic energy into heat.  Exceptions are the work of \cite{dd08}, \cite{lg10}, \cite{dd10} and \cite{dd12}, which include proper shock capturing.  The simulations of \cite{lg10} are 2D and require confirmation by 3D simulations due to the global nature of flow features in the atmospheres of close-in, irradiated exoplanets.  Although their simulations are 3D in nature, \cite{dd08} and \cite{dd10} employed a non-global grid in an attempt to circumvent the ``pole problem" \citep{st12}, where the convergence of meridians at the poles results in vanishing time steps for the simulations.  \cite{dd12} employed a partial solution to the pole problem that violates causality.  Detailed comparisons between the work of \cite{dd12} and the insight gained from the present study require future 3D simulations that solve the pole problem and include shock heating in a self-consistent, energy-conserving manner.

\cite{php12} examined the effects of varying stellar irradiation on the atmospheric circulation.  At locations where the Mach number exceeds unity, they assumed a shock to form and dissipate a fraction $4 (\gamma-1)/(\gamma+1)^2$ of the kinetic energy into heat.  In light of our present results, such an approach produces an over-estimation of the shock heating.  Nevertheless, it remains likely that shocks do not penetrate deeply enough into the atmosphere to affect the evolution of the exoplanet in a significant manner.

\subsection{Observational Consequences}

We have demonstrated that shocks are expected to be present near the infrared photospheres of sufficiently irradiated exoplanets.  The main effect of shocks is to convert some fraction $f_s$ of the local kinetic energy of the atmosphere into heat.  We now wish to derive an expression for $f_s$ and also estimate the ratio $r_s$ of the flux of shock heating to the incident stellar flux.

We start from the shock jump condition for temperature \citep{lr57},
\begin{equation}
\frac{T^\prime}{T} = 1 + \frac{2\left(\gamma-1\right)}{\left(\gamma+1\right)^2} \left( \frac{{\cal M}^2 - 1}{{\cal M}^2} \right) \left( \gamma {\cal M}^2 + 1 \right).
\end{equation}
Using $\gamma {\cal M}^2 = m v^2/kT$ and recognizing that $f_s = 2kT^\prime/mv^2$ and $f_{s_0} = 2 k T/mv^2$, we obtain
\begin{equation}
f_s = f_{s_0} + \frac{4\left( \gamma - 1 \right)}{\left( \gamma + 1 \right)^2} \left[ 1 - \frac{\left( \gamma -1 \right)}{\gamma {\cal M}^2} - \frac{1}{\gamma {\cal M}^4} \right],
\end{equation}
where $f_{s_0}$ is the fraction of thermal to kinetic energy in the pre-shock flow.  For example, if $f_{s_0} = 0$ and ${\cal M}=2$ then we obtain $f_s \approx 0.25$.  The ratio of this shock heating to the incident stellar energy is
\begin{equation}
r_s \sim \frac{f_s \sigma_{\rm SB} k_{\rm B} v^2 T^3}{2 {\cal F}_0 \kappa c_P^2 m} \exp{\left( \tau_{\rm S} \right)},
\end{equation}
where $\sigma_{\rm SB}$ denotes the Stefan-Boltzmann constant, $k_{\rm B}$ the Boltzmann constant, $\tau_{\rm S} \sim 1$ the shortwave optical depth, ${\cal F}_0$ the top-of-the-atmosphere stellar flux, $\kappa = 2/(2 + n_{\rm dof})$ the adiabatic coefficient, $c_P$ the specific heat capacity at constant pressure and $m$ the mean molecular mass.  For Model H, we estimate that $r_s \sim 0.1$ at locations where shock formation occurs.

Shock heating functions as an additional form of drag, which acts to reduce the efficiency of heat redistribution from the dayside to the nightside hemisphere of a exoplanet.  Consequently, the day-night flux contrast is expected to increase, while the shift of the peak of the phase curve from the substellar point is expected to decrease.  Simulations that fail to capture shock heating will thus \emph{over-estimate} the peak offset.  

\vspace{0.2in}
\textit{This work was inspired by a dinner conversation with Ray Pierrehumbert during the Exoclimes II conference at the Aspen Center for Physics in early 2012.  KH thanks the anonymous referee for constructive comments that improved the veracity and clarity of the manuscript.  KH acknowledges generous support by the Zwicky Prize Fellowship of ETH Z\"{u}rich and the Swiss-based MERAC Foundation.}


\label{lastpage}

\end{document}